# Superconductivity in Ternary Germanide ScPdGe and Silicide ScPdSi

Yusaku Shinoda[1], Yoshihiko Okamoto[1,2*], Youichi Yamakawa[3], Hiroshi Takatsu[4], Hiroshi Kageyama[4], Daigorou Hirai[1], and Koshi Takenaka[1]

[1]*Department of Applied Physics, Nagoya University, Nagoya 464-8603, Japan*

[2]*Institute for Solid State Physics, University of Tokyo, Kashiwa 277-8581, Japan*

[3]*Department of Physics, Nagoya University, Nagoya 464-8602, Japan*

[4]*Graduate School of Engineering, Kyoto University, Kyoto 615-8510, Japan*

The electronic properties of ScPdGe and ScPdSi, crystallizing in the hexagonal ZrNiAl and orthorhombic TiNiSi structures, respectively, are investigated. ScPdGe and ScPdSi are found to show bulk superconductivity below 0.9 and 1.7 K, respectively, based on electrical resistivity and heat capacity data measured using synthesized polycrystalline samples. First-principles calculations indicate the presence of large contributions of Sc 3*d* and Pd 4*d* electrons at the Fermi energy in both materials. The electronic properties and electronic states of these materials are discussed in comparison with those of several superconductors containing scandium and a 4*d* transition metal element.

Superconductivity in 3*d* electron systems, such as high-$T_c$ superconductivity in cuprates and iron arsenides, is one of the central issues in condensed matter physics [1,2]. In these systems, the various features of 3*d* electrons, such as strong electron correlation and spin and orbital fluctuations, lead to various unconventional superconductivity. Many unconventional superconductors have also been discovered in 4*d* and 5*d* electron systems, where the features of 4*d* and 5*d* electrons, such as strong spin–orbit coupling and orbital degrees of freedom, play a role. Parity-mixing superconductivity in $Li_2Pt_3B$ and the Fulde–Ferrell–Larkin–Ovchinnikov state in $Sr_2RuO_4$ are fascinating examples [3,4]. However, most of the outstanding *d*-electron superconductors discovered thus far contain one transition metal element. There are few *d*-electron superconductors containing both 3*d* and 4*d*/5*d* transition metal elements that exhibit remarkable superconducting properties. The interplay between 3*d* and 4*d*/5*d* electrons may lead to unprecedented electronic properties in such materials, and the combination of multiple transition metals can provide a wider variety of materials than those with one transition metal, suggesting that the materials with both 3*d* and 4*d*/5*d* transition metals are worth exploring for new superconductors.

Ternary compounds of Sc, M, and X, where M is a 4*d*/5*d* transition metal element and X is a more electronegative element than Sc and M, are promising as *d*-electron superconductors containing both 3*d* and 4*d*/5*d* transition metal elements. Although Sc is a first period transition metal, in which 3*d* electrons can be involved in the electronic prop-erties, it has a considerably larger ionic radius than those of other 3*d* transition metals. Materials containing multiple transition metals often have large site disorder due to the occupation of one crystallographic site by them. However, in the Sc–M–X ternary compounds, Sc and M atoms almost always occupy different crystallographic sites due to the large difference in ionic radii, resulting in the crystal structures without site disorder. This is advantageous for realizing unconventional superconductivity, where Cooper pairs are easily broken down by disorder. In fact, several superconductors have been discovered in the Sc–M–X ternary compounds, such as ScIrP and $Sc_6OsTe_2$, in which both the Sc 3*d* and Ir/Os 5*d* electrons are strongly involved in the superconducting properties [5,6].

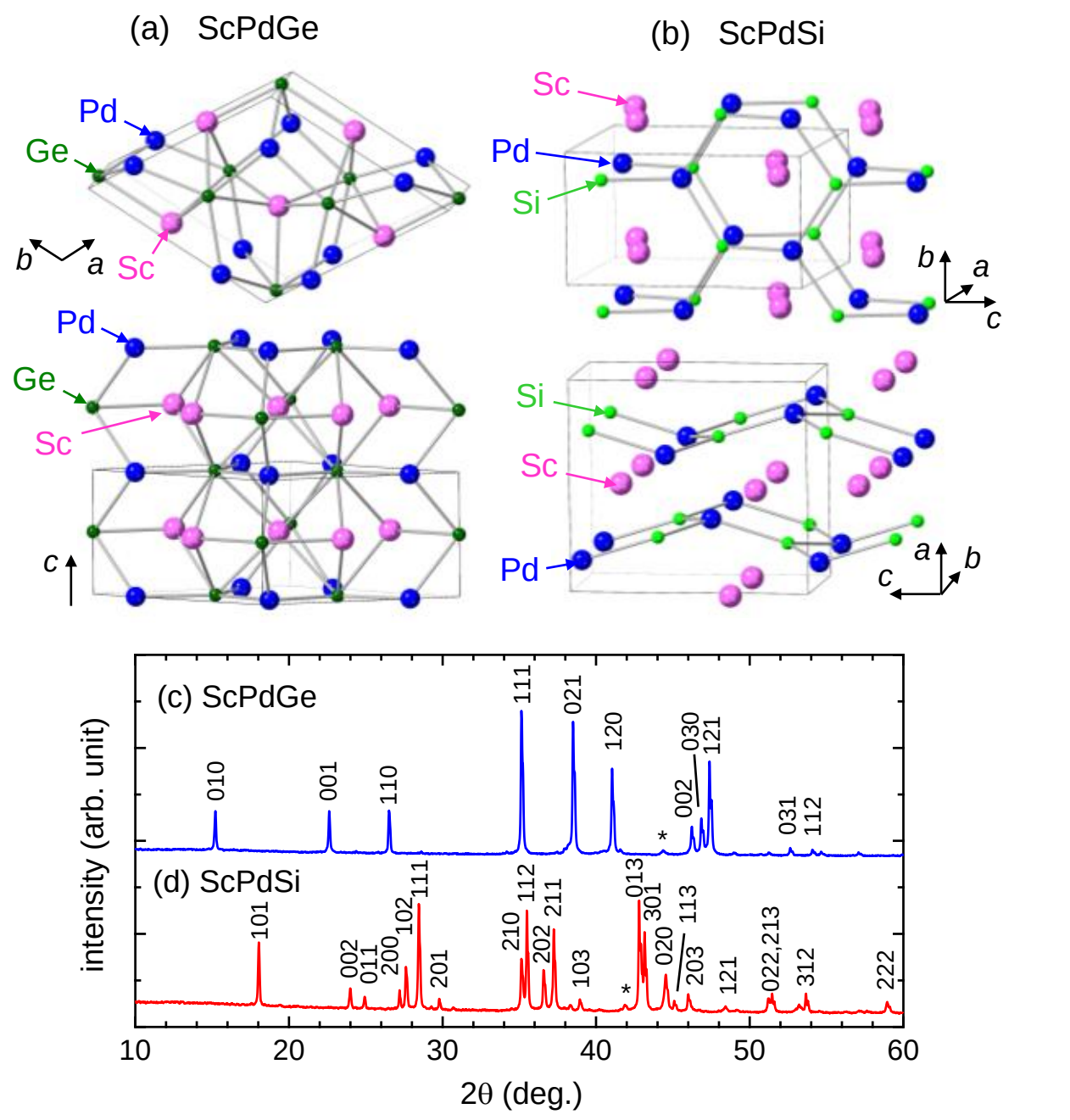

Figure 1. Crystal structures of (a) ScPdGe and (b) ScPdSi. The thin solid lines show the unit cell. Powder X-ray diffraction patterns of polycrystalline samples of (c) ScPdGe and (d) ScPdSi measured at room temperature. The stars indicate the diffraction peaks of an unknown impurity. Peak indices are given using a hexagonal unit cell with lattice constants of $a$ = 6.7183(8) Å and $c$ = 3.9278(5) Å for ScPdGe, and an orthorhombic unit cell with $a$ = 6.558(2) Å, $b$ = 4.0696(8) Å, and $c$ = 7.416(2) Å for ScPdSi.

In this letter, we report that ScPdGe and ScPdSi show bulk superconductivity below 0.9 and 1.7 K, respectively. There have been no reports of the electronic properties of these materials in previous studies. ScPdGe crystallizes in the ZrNiAl structure, which is an ordered $Fe_2P$ type, as shown in Fig. 1(a) [7,8]. This crystal structure is noncentrosymmetric with a hexagonal space group of *P*−62*m* [7,8]. In contrast, ScPdSi crystallizes in the TiNiAl structure, which is an ordered $Co_2P$ type, as shown in Fig. 1(b) [7,9]. This crystal structure is centrosymmetric with an orthorhombic space group of *Pnma* and consists of alternately stacking of Sc atoms and a distorted honeycomb net made of Pd and Si atoms. In ZrNiAl-type Sc compounds, ScIrP was reported to show bulk superconductivity below $T_c$ = 3.4 K [5]. Theoretical calculations suggested that this superconductivity is phonon-mediated *s*-wave superconductivity [10]. ScRhP was also reported to exhibit a superconducting transition with $T_c$ = 1−2 K, which has not been confirmed to be bulk superconductivity [11]. ScPdGe has the same number of valence electrons as ScIrP and ScRhP, because a Pd atom has one more valence electron than a Rh or Ir atom and a Ge atom has one less valence electron than a P atom. In TiNiSi-type Sc compounds, ScRuSi was reported to show a bulk superconducting transition at 3.1 K [12].

Polycrystalline samples of ScPdGe and ScPdSi were synthesized by the arc-melting method. Equimolar Sc chips, Pd powder, and Ge or Si powder were weighed. The Pd and Ge/Si powders were then mixed and pressed into a pellet. The Sc chips and pellet were put on a water-cooled copper hearth and arc-melted under an Ar atmosphere. The obtained buttons were arc melted several times to promote homogenization. The synthesized samples were characterized by powder X-ray diffraction employing Cu Kα radiation at room temperature using a MiniFlex diffractometer (RIGAKU). As shown in Figs. 1(c) and 1(d), all the diffraction peaks, except for small peaks from unknown impurity phases, were indexed based on a hexagonal unit cell with $a$ = 6.7183(8) Å and $c$ = 3.9278(5) Å for ScPdGe and an orthorhombic unit cell with $a$ = 6.558(2) Å, $b$ = 4.0696(8) Å, and $c$ = 7.416(2) Å for ScPdSi. These lattice constants were almost the same as those reported in previous studies [7–9], indicating that the obtained samples were almost single phases of the ZrNiAl-type ScPdGe and the TiNiSi-type ScPdSi. Electrical resistivity and heat capacity measurements were performed using a Physical Property Measurement System (Quantum Design). Electrical resistivity down to 0.1 K was measured using an adiabatic demagnetization refrigerator. Heat capacity down to 0.6 K was measured using a $^3$He refrigerator. Electronic structure calculations were performed using the WIEN2k package in the framework of density functional theory (DFT) based on the full-potential linearized augmented plane wave (FP-LAPW) method [13]. We used the Perdew−Burke−Ernzerhof (PBE) functional, which is one of the most widely used generalized gradient approximation (GGA) based exchange-correlation functionals. Experimentally determined structural parameters were used for the calculations [8,9].

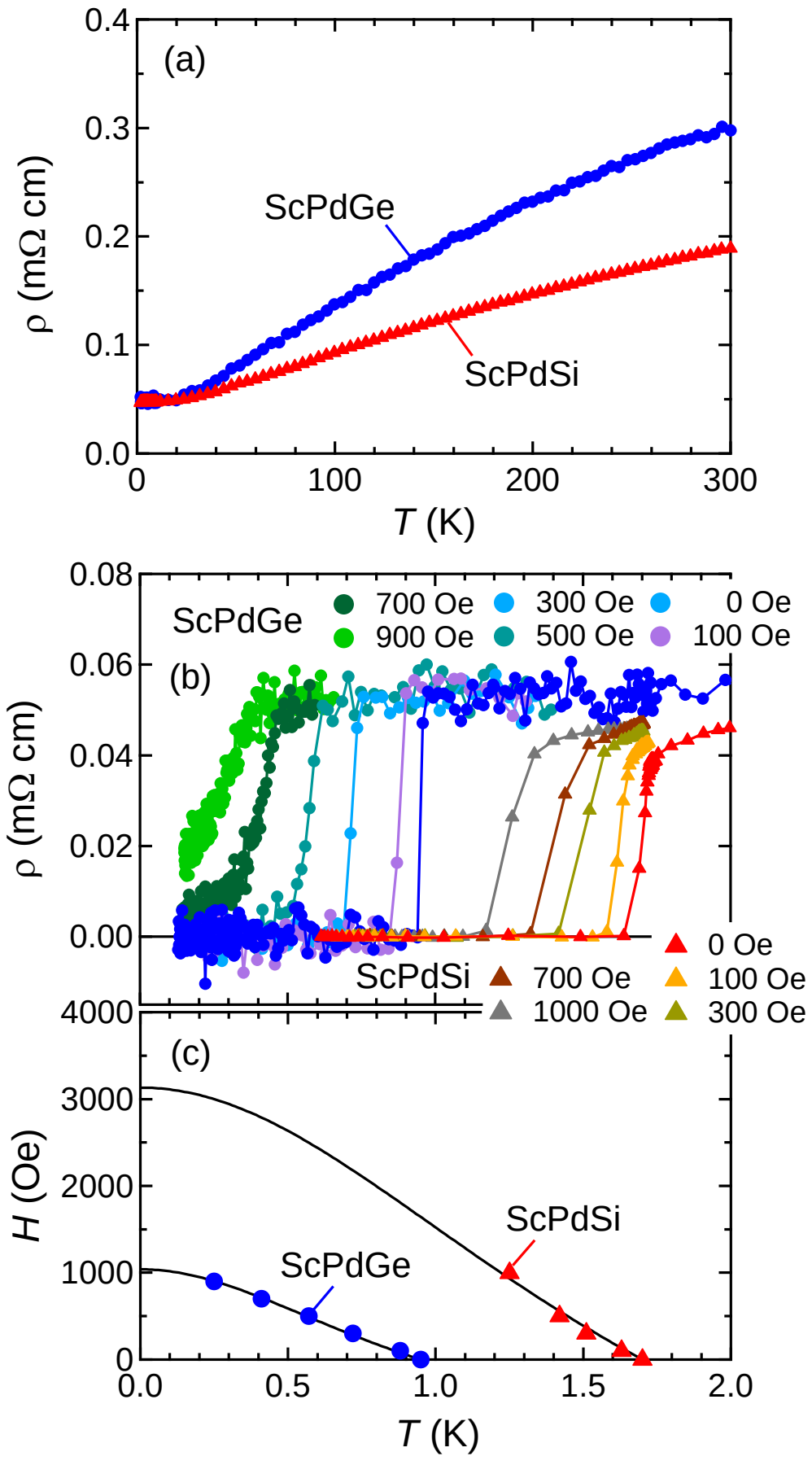

Figure 2. (a) Temperature dependence of electrical resistivity of polycrystalline samples of ScPdGe and ScPdSi. (b) Electrical resistivity of ScPdGe and ScPdSi at low temperatures under various magnetic fields. (c) Magnetic fields versus midpoint temperatures in the resistivity drop for ScPdGe and ScPdSi. The solid curves show a fit to the Ginzburg–Landau formula.

Temperature dependences of the electrical resistivity ρ of the ScPdGe and ScPdSi polycrystalline samples are shown in Fig. 2(a). Both samples showed metallic behavior with ρ decreasing with decreasing temperature. The residual resis-tivity ratio, RRR = $\rho_{300K}/\rho_0$, where $\rho_{300K}$ is ρ at 300 K and $\rho_0$ is the residual resistivity, is equal to 6 and 4 for ScPdGe and ScPdSi, respectively. When the temperature was further decreased, the ρ of both ScPdGe and ScPdSi rapidly decreased to zero, as shown in Fig. 2(b). These zero resistivity temperatures were suppressed by applying magnetic fields, indicative of the superconducting transition. The onset temperature of the resistivity drop and the zero resistivity temperatures were 0.97 and 0.94 K for ScPdGe and 1.73 and 1.68 K for ScPdSi, respectively.

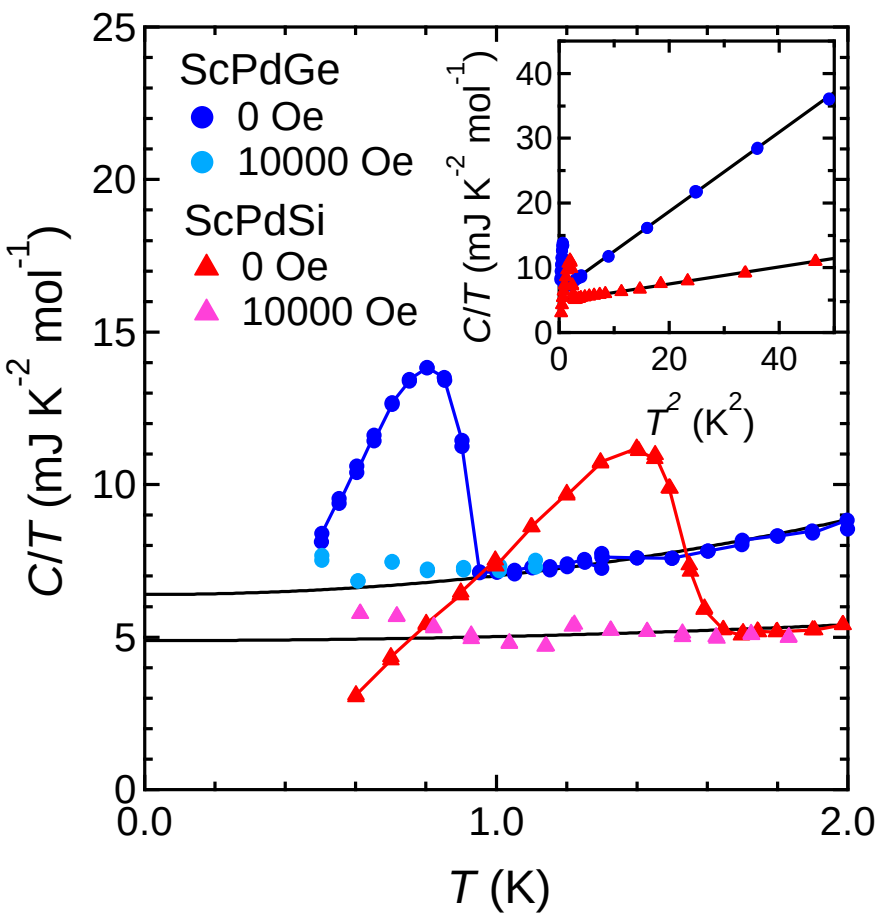

Figure 3. Temperature dependence of heat capacity divided by temperature as a function of $T$ (main panel) and $T^2$ (inset) measured at magnetic fields of 0 and 10000 Oe. The solid lines (inset) and curves (main panel) show a fit to the equation $C/T = AT^2 + \gamma$ for the 1−5 K data for ScPdGe and 1.9−6 K data for ScPdSi measured at 0 Oe.

Figure 3 shows the temperature dependence of the heat capacity divided by temperature, $C/T$, of the ScPdGe and ScPdSi polycrystalline samples. The values of $C/T$ strongly increased below 0.9 K for ScPdGe and 1.7 K for ScPdSi, followed by clear peaks, indicating that a phase transition occurs in both samples. These temperatures correspond to the zero resistivity temperatures, which clearly indicate that the superconducting transition in both samples is a bulk superconducting transition. The emergence of a peak in the $C/T$ data of ScPdGe and ScPdSi is in contrast to the case of isoelectric ScRhP, where only a gradual increase was observed at around $T_c$ [11], supporting the high quality of the ScPdGe and ScPdSi samples. The peaks are completely suppressed by applying a magnetic field of 10000 Oe, which is consistent with the magnetic field effect on ρ shown in Fig. 2(b). Considering the zero-resistivity temperatures of 0.94 and 1.68 K and heat capacity jumps of 0.9 and 1.7 K for ScPdGe and ScPdSi, respectively, the critical temperatures were determined as $T_c = 0.9$ and 1.7 K for ScPdGe and ScPdSi, respectively.

We then considered the superconducting properties of ScPdGe and ScPdSi based on the electrical resistivity under magnetic fields and heat capacity data. As shown in Fig. 2(b), the superconducting transitions in ScPdGe and ScPdSi were not completely suppressed by applying a magnetic field of 900 Oe for ScPdGe and 1000 Oe for ScPdSi, suggesting that they are type-II superconductors. Figure 2(c) shows the plots of magnetic fields versus midpoint temperatures in the resistivity drop for ScPdGe and ScPdSi. Each plot was fitted to the Ginzburg–Landau (GL) formula $H_{c2}(T) = H_{c2}(0)[1 - (T/T_c)^2]/[1 + (T/T_c)^2]$, assuming type-II superconductivity, yielding $H_{c2}(0) = 1.04(1) \times 10^3$ and $3.1(2) \times 10^3$ Oe and the GL coherence length $\xi_{GL}$ of 56 and 33 nm for ScPdGe and ScPdSi, respectively.

The solid curves in the main panel of Fig. 3 shows the fitting results to the equation $C/T = AT^2 + \gamma$, where $A$ and γ represent the coefficient of the $T^3$ term of the lattice heat capacity and the Sommerfeld coefficient, respectively. The $A$ and γ for ScPdGe were estimated to be $A$ = 0.614(3) mJ $K^{-4}$ $mol^{-1}$ and γ = 6.40(3) mJ $K^{-2}$ $mol^{-1}$ by the linear fit of the $C/T$ versus $T^2$ plot between 1 and 5 K in the normal state, as shown in the inset of Fig. 3. The similar fit between 1.9 and 6 K for the ScPdSi data yielded $A$ = 0.131(2) mJ $K^{-4}$ $mol^{-1}$ and γ = 4.89(3) mJ $K^{-2}$ $mol^{-1}$. These $A$ values give Debye temperatures of $\theta_D$ = 211 and 354 K for ScPdGe and ScPdSi, respectively. As shown in the main panel of Fig. 3, the entropy balance between the superconducting and normal states seems to be well maintained for each material, although quantitative evaluation is not possible due to the absence of experimental data far below $T_c$. The magnitudes of the jump at $T_c$ were estimated to be $\Delta C/\gamma T_c$ = 1.1 for ScPdGe and 1.3 for ScPdSi. Although these values are slightly smaller than the weak-coupling limit value of 1.43 for a conventional BCS superconductivity, this difference may be due to residual γ caused by the imperfections and defects in the polycrystalline samples. The superconductivity of ScPdGe and ScPdSi is most likely the weak-coupling BCS superconductivity.

Next, we discuss the electronic states of ScPdGe and ScPdSi compared with those of ScRhP. These three materials are isoelectric and ScPdGe and ScRhP are isostructural. Figure 4(a) shows the electronic band structure and electronic density of states (DOS), $D(E)$, of ScPdGe, calculated with spin–orbit coupling. The band structure of ScPdGe near $E_F$ is similar to that of ScRhP [11,14]. There are complex Fermi surfaces around $k_z \sim 0$ and the band dispersions cross $E = E_F$ near the midpoint of the Γ–A line, i.e., $k_z \sim \pi/2c$. At $k_z \sim \pi/c$, there is another Fermi surface surrounding the A point. The Sc 3$d$ and Pd 4$d$ orbitals greatly contribute to the electronic states at $E_F$, whereas the contribution of Ge is small. In contrast, the structures of DOS near $E_F$ are considerably different between ScPdGe and ScRhP. In ScRhP, $E_F$ is located at the large peak of DOS. The DOS at $E_F$ is $D(E_F)$ = 9.4 $eV^{-1}$, yielding a calculated Sommerfeld coefficient of $\gamma_{band}$ = 7.4 mJ $K^{-2}$ $mol^{-1}$ [11]. In contrast, there is only a small peak in DOS of ScPdGe at around $E_F$, giving rise to smaller $D(E_F)$ of 6.3 $eV^{-1}$ and $\gamma_{band}$ of 4.9 mJ $K^{-2}$ $mol^{-1}$ than those of ScRhP. The DOS structure of ScPdGe near $E_F$ is rather similar to that of ScIrP [$D(E_F) = 5.2$ eV$^{-1}$ and $\gamma_{band} = 4.05$ mJ K$^{-2}$ mol$^{-1}$], which is a 5*d* analogue of ScRhP [11]. The difference between ScPdGe and ScRhP is due to the lower energy level of Pd 4*d* orbitals than that of Rh 4*d* orbitals, resulting in the smaller contributions of Pd 4*d* electrons at $E_F$ than those of Rh 4*d* electrons. Nevertheless, ScPdGe and ScRhP have similar experimental $\gamma$ ($\gamma$ = 6.4 and 7.5 mJ K$^{-2}$ mol$^{-1}$, respectively) and $T_c$ (the zero-resistivity temperature of ScRhP is 1 K) [11]. In contrast, the critical magnetic field of ScPdGe is much lower than that in ScRhP ($H_{c2}(0) = 5000$ Oe for ScRhP). This may be related to the smaller contribution of Pd 4*d* electrons of ScPdGe at $E_F$ than that of Rh 4*d* electrons of ScRhP, which is the largest difference of electronic states between ScPdGe and ScRhP.

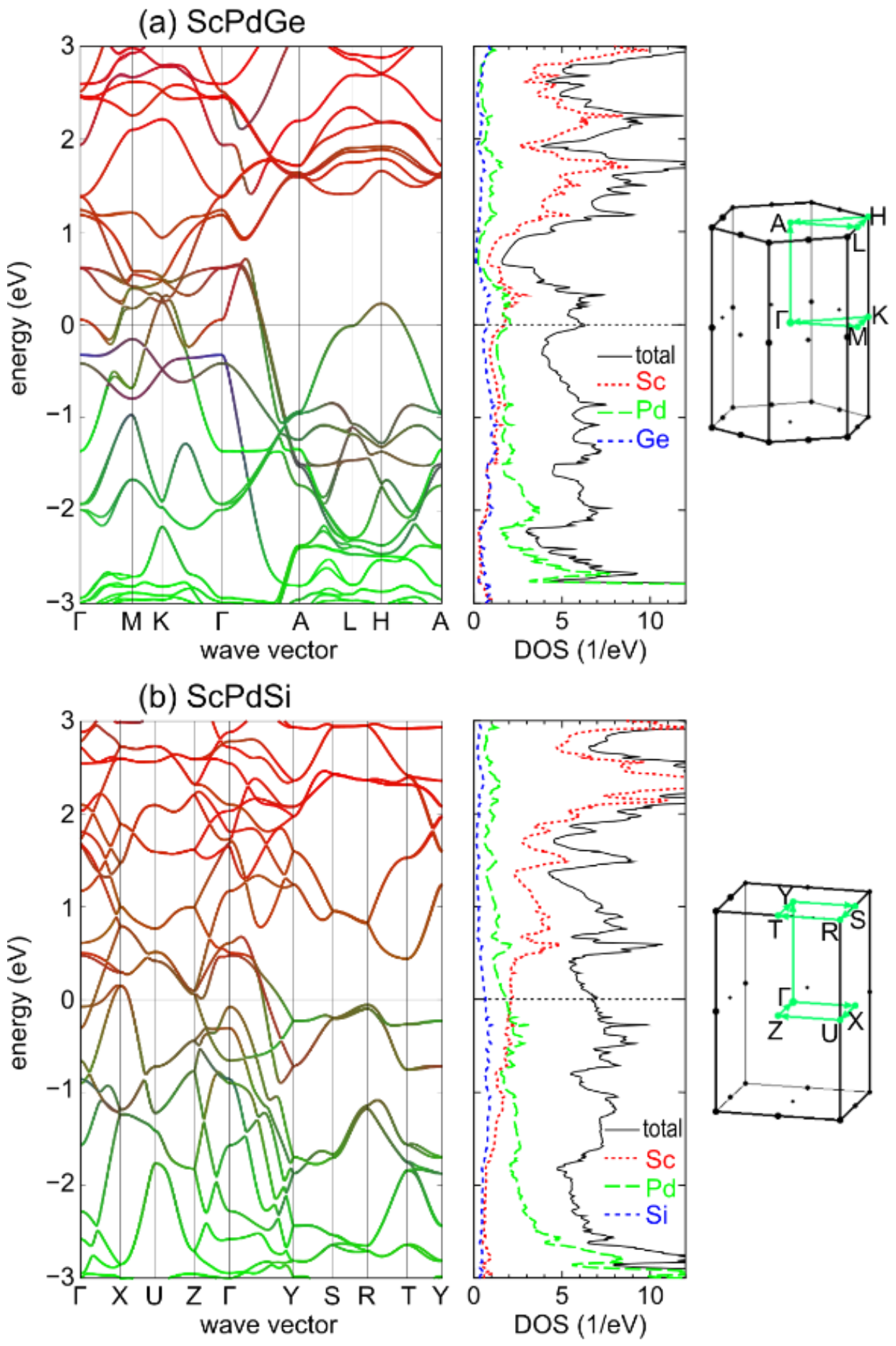

Figure 4. Electronic states of (a) ScPdGe and (b) ScPdSi with spin–orbit coupling. Electronic structures (left), total and partial DOS (center), and the first Brillouin zone (right) are shown. The green, red, and blue colors in the electronic structures represent the contributions of Sc, Pd, and Si/Ge, respectively. The Fermi energy is set to 0 eV.

Figure 4(b) shows the electronic structure and DOS of ScPdSi calculated with spin–orbit coupling. In ScPdSi, there are large contributions of Sc 3*d* and Pd 4*d* orbitals at $E_F$ and that of Si is small, similar to the case of ScPdGe. The DOS at $E_F$ in ScPdSi is $D(E_F) = 6.6$ eV$^{-1}$, yielding $\gamma_{band} = 3.9$ mJ K$^{-2}$ mol$^{-1}$, which is slightly smaller than the experimental $\gamma = 4.9$ mJ K$^{-2}$ mol$^{-1}$. Although this $\gamma$ value is smaller than $\gamma = 6.4$ mJ K$^{-2}$ mol$^{-1}$ for ScPdGe, ScPdSi exhibited approximately twice higher $T_c$ than that of ScPdGe. As discussed above, the heat capacity data suggested that the superconductivity in both compounds are phonon-mediated weak-coupling BCS superconductivity. Therefore, the higher $T_c$ in ScPdSi than that in ScPdGe is most likely due to the fact that ScPdSi contains silicon, which is a lighter element than germanium. In fact, ScPdSi showed a higher $\theta_D$ of 354 K than $\theta_D = 211$ K in ScPdGe. This $\theta_D$ of ScPdSi is comparable to $\theta_D = 388$ K in isostructural ScRuSi with $T_c = 3.1$ K [12]. Thus, ScPdGe and ScPdSi exhibited considerably different $T_c$ values probably due to the difference between Ge and Si. However, these superconductors are expected to serve as good model compounds for discussing the influence of spatial inversion symmetry on *d*-electron superconductors containing both 3*d* and 4*d* transition metal elements, which will be elucidated in future experiments using single crystals.

In summary, ScPdGe and ScPdSi, both comprising 3*d* and 4*d* transition metal elements, are found to be bulk superconductors. ScPdGe crystallizes in the hexagonal ZrNiAl structure without space inversion symmetry, whereas ScPdSi crystallizes in the TiNiSi structure, which is centrosymmetric and orthorhombic. The electrical resistivity and heat capacity data of the synthesized polycrystalline samples indicated that ScPdGe and ScPdSi showed bulk superconducting transition at $T_c = 0.9$ and 1.7 K, respectively. The heat capacity data can be explained within the weak-coupling BCS theory. The first principles calculations on ScPdGe and ScPdSi indicated that both compounds have considerable contributions from Sc 3*d* and Pd 4*d* electrons to the electronic states at $E_F$.

**Acknowledgments**

This work was partly carried out at the Materials Design and Characterization Laboratory under the Visiting Researcher Program of the Institute for Solid State Physics, University of Tokyo, and supported by JSPS KAKENHI (Grant Numbers: 18H04314, 19H05823, and 19K21846).

*e-mail: yokamoto@issp.u-tokyo.ac.jp